**Title**

Multilayer Epitaxial Graphene Formed by Pyrolysis of Polycrystalline Silicon-Carbide Grown on C-Plane Sapphire Substrates

**Authors**

Timothy J. McArdle, Jack O. Chu, Yu Zhu, Zihong Liu, Mahadevaiyer Krishnan, Christopher M. Breslin, Christos Dimitrakopoulos, Robert Wisnieff, and Alfred Grill

**Affiliation**

IBM T. J. Watson Research Center, Yorktown Heights, NY 10598

**Abstract**

We use ultra-high vacuum chemical vapor deposition to grow polycrystalline silicon carbide (SiC) on c-plane sapphire wafers which are then annealed between 1250 and 1450°C in vacuum to create epitaxial multilayer graphene (MLG). Despite the surface roughness and small domain size of the polycrystalline SiC, a conformal MLG film is formed. By planarizing the SiC prior to graphene growth, a reduction of the Raman defect band is observed in the final MLG. The graphene formed on polished SiC films also demonstrates significantly more ordered layer-by-layer growth and increased carrier mobility for the same carrier density as the non-polished samples.

**Article Text**

Since the isolation of micron-sized exfoliated graphene flakes by Novoselov and Geim in 2004[1], a considerable amount of research has been devoted to scaling up the production of high-quality graphene films to a size suitable for use in the mass production of electronic devices. Unfortunately, all of the best studied production methods have unique limitations which make integration with existing silicon technology and processing techniques difficult.

Arguably the most promising candidate technique, epitaxial growth of graphene by removal of surface silicon from single-crystal silicon carbide (SiC) wafers, has produced highly uniform mono- and few-layer graphene samples with high carrier Hall mobility[2,3,4] and transistors capable of radio frequency (RF) cutoff frequencies up to 170 GHz.[5] However, this epitaxial technique is limited by the simple fact that no SiC wafers larger than 5" in diameter are currently available. Growth of graphene on metal foil by chemical vapor deposition (CVD) produces films that are self-limited to a monolayer of graphene and can be grown in almost unlimited sizes (up to 30" wide, continuous rolls have been reported[6]). But the CVD grown graphene consists of relatively small graphene domains which appear to limit the intrinsic Hall mobility and final device performance[7] and must be transferred from the conducting foil to an insulating substrate prior to processing and integration into circuits. More recently, synthesis of graphene from silicon carbide films grown on silicon substrates has been demonstrated.[8,9,10] However, this technique is limited to process temperatures well below the 1410°C melting point of silicon and, even at lower process temperatures, requires thick (>500 nm) layers of SiC to ensure that the Si does not evaporate through the SiC film. While graphene may be formed at such low temperatures in the correct environment, many epitaxial graphene studies have shown marked improvements in film quality by annealing at higher temperatures in gas environments.[2,11,12]

In this letter, we present results for multi-layer graphene (MLG) films produced from polycrystalline SiC grown on c-plane sapphire substrates. Using sapphire as the substrate has certain advantages, such as the availability of large (up to 12") diameter wafers, excellent performance in RF and radiation hardened applications, and the ability to withstand processing temperatures much higher than those used for the formation of graphene. Because of the roughness of the non-planarized, polycrystalline SiC surface, the MLG films actually appear to be mixtures consisting of graphene and carbon bundles as demonstrated by Raman, cross-sectional transmission electron microscopy (TEM), and electrical characterization. However, planarization of the SiC by chemical-mechanical polishing (CMP) prior to the formation of the MLG dramatically improves the quality of the graphene obtained and stops any formation of carbon bundles.

The SiC films were grown in an ultra-high vacuum (UHV) chemical vapor deposition (CVD) system at between 900 and 1100°C, on either 2" or 4" diameter, epitaxial-grade c-plane sapphire wafers. The SiC was grown in two stages; first a thin buffer layer was deposited at 950°C, then a thick overlayer was deposited at 1050°C. The final thickness of the films ranged from 80 nm to more than 700 nm. Some of the 4" diameter SiC films were planarized by CMP before being annealed to form graphene.

TEM images and an x-ray diffraction (XRD) spectrum from a typical 2" diameter, 84 nm thick SiC on sapphire sample are presented in Figure 1. The XRD spectrum [Fig. 1(a)] shows that the SiC film is polycrystalline with the bulk of the crystals having the 3C polytype, while the electron diffraction images [Fig. 1(b)] demonstrate that the SiC domains share the same orientation as the sapphire substrate. The TEM image in Fig. 1(c) shows conformal coverage of the sapphire by small, crystalline grains of SiC about 40 to 80 nm wide. Topographic atomic force microscope (AFM) images (not shown for this sample) confirm that the surface is continuous but rough, with a root-mean square (RMS) roughness of 3.2 nm.

Formation of the graphene layers took place in a second UHV processing chamber which has a base pressure of $6.0 \times 10^{-10}$ Torr and is equipped with two parallel graphite-filament, pyrolytic boron nitride (PBN) coated heater stages. A standard tungsten/rhenium thermocouple was used to monitor the process temperature. For this study, samples were first outgassed for 5 minutes at 820°C in vacuum and then cleaned by flowing 20% disilane in helium for 10 minutes at $1.0 \times 10^{-6}$ Torr with the sample still held at 820°C. Next the disilane flow was stopped and samples were annealed for 10 minutes at temperatures ranging from 1250 to 1450°C in ambient vacuum environment. The total pressure during the final annealing step was dominated by nitrogen outgassing from the PBN heater coating and ranged from $1.0 \times 10^{-7}$ to $1.0 \times 10^{-6}$ Torr.

A Raman spectrum from a typical sample [Fig. 2(a)] confirms the presence of graphene after annealing at 1420°C for 10 minutes in vacuum and appears similar to spectra reported by Ogawa *et. al.*.[13] The strong D-band (1358 cm$^{-1}$) peak in the spectrum indicates the presence of defects in the graphene film[14] and is most likely caused by the boundaries of many small graphene domains. The TEM images [Fig. 2(b)] show that the film has formed a conformal layer over the SiC despite its surface roughness and grain boundaries. From the TEM image, it is also clear that the film is thick—ranging from 3.9 to 21 nm—and highly non-uniform. Higher magnification TEM (not shown) suggests that the growth of the MLG is controlled by the facets of the individual SiC crystallites and can occur in large, vertical bundles similar to carbon nanotubes (CNT). This is not too surprising, as it is possible to grow ordered arrays of CNT on commercial, c-face SiC wafers under certain conditions.[15] We attribute the mixed growth mode of multi-layer graphene and nanotubes observed here to the rough surface and polycrystalline nature of these SiC films. The high surface energy of the boundaries between crystallites creates a natural place for Si to escape during the annealing stage, allowing the MLG and CNT film to form more quickly and non-uniformly than it does on atomically flat, single-crystal SiC wafers.

Figure 3 compares AFM and Raman of samples from two batches of 4" SiC on sapphire wafers. The first batch of SiC films was grown under the same conditions and to the same thickness as the 2" sample shown in Figure 1. The second batch of SiC wafers was also grown under similar conditions, but to a thicker SiC film (>700 nm) suitable for planarization by CMP. Topographic AFM of an as-grown thin SiC film from the first batch is shown in Fig. 3(a) and AFM of the post CMP surface of the thicker SiC from the second batch is shown in Fig. 3(d). Although the CMP process has not yet been optimized for this material and the result is not a fully planarized film, the RMS surface roughness of the thick SiC film has been reduced from 14 nm to 4.5 nm, and—more importantly—relatively large, flat, and connected areas have been formed which are not found on any of the non-CMP samples.

After the formation of graphene by annealing for 10 minutes at 1300°C in vacuum, the surface morphology of the non-CMP sample appears unchanged according to the AFM [Fig. 3(b)] while the CMP sample has roughened slightly (to 4.8 nm) but retained much of the flat area structure that the non-CMP film lacks [Fig. 3(e)]. The effect of the different surfaces is readily seen in the Raman spectra of the two samples [Figures 3(c) and 3(f)]. Because the D to G peak intensity ratio is inversely proportional to the graphene domain size,[12, 14] comparing the peak intensity ratios from the two samples provides a relative estimate of the quality of the films. The D to G ratio for the non-CMP film is 0.84, while the CMP sample has a significantly lower ratio of 0.52, which suggests that it consists of larger graphene domains.

From the cross sectional TEM images of the CMP film shown in Figures 3(g) and 3(h), it is also clear that the graphene on the CMP SiC has grown in a more controlled, layer by layer manner, than the non-CMP film. Instead of a thick film of material growing in bundles and along multiple crystalline directions, the CMP film produced 15 to 20 layers of graphene (with layer spacing of 3.4 Å) constrained to be parallel to the flattened SiC grains and comparable to films normally observed on c-face SiC. Even in areas where the SiC domains were not entirely flattened or where a grain has a rotated orientation, layers of graphene can still be resolved and there is no evidence of any CNT growth. Low energy electron diffraction (LEED) taken at 76 eV on a sample similar to the one shown in Figure 3 reveals diffuse first order graphene spots superimposed on a faint ring. The LEED spots are diffuse because of the MLG film's surface roughness, but their presence indicates that many of the graphene domains are aligned along a primary direction. The faint ring is caused by the remaining polycrystalline domains which are rotated randomly. Similar rotational stacking faults between graphene layers are also observed in graphene grown on commercial c-face SiC.[16] The number of graphene layers formed on the CMP SiC films can be controlled through the anneal temperature, ambient pressure, SiC surface condition, and anneal time. A sample prepared identically to the one shown in Fig. 3, but annealed at slightly lower heater power, formed 10 to 12 layers of graphene and had a D to G peak intensity ratio of 0.37.

Large area (160 μm x 200 μm channel) Hall bars were fabricated on the two samples in Fig. 3 using standard optical photolithography to determine the carrier mobility, carrier density, and sheet resistance. The details of the device geometry and processing have been described elsewhere.[2] Devices were tested at room temperature in a high vacuum Hall probe system after being baked *in situ* overnight at 400 K. The source current was 100 μA and the magnetic field was swept ±1.0 Tesla. For the non-CMP film, the average sheet resistance across multiple devices was 5.18 ± 0.45 kOhm/square, with a carrier density of 1.22 ± 0.13 x $10^{13}$ $cm^{-2}$ and a mobility of 100.6 ± 3.8 $cm^2$/Vs. For the CMP film, the average sheet resistance across multiple devices was 3.12 ± 0.12 kOhm/square, with a carrier density of 1.13 ± 0.03 x $10^{13}$ $cm^{-2}$ and a mobility of 178.4 ± 3.6 $cm^2$/Vs. For similar carrier densities, the mobility of the MLG on the CMP film is 77% higher than the non-CMP film. This higher carrier mobility observed in the CMP film is a direct consequence of the film's lower sheet resistance, which we attribute to the more ordered growth, lack of CNT-like bundles, and larger domain size of the graphene layers.

In summary, we have formed multi-layer graphene by silicon sublimation from polycrystalline SiC films grown on sapphire substrates up to 4" in diameter. Additionally, we demonstrate that the quality of the epitaxial graphene may be greatly improved by polishing the SiC prior to annealing. Further optimization of the CMP process to produce flatter SiC starting surfaces, as well as better understanding and control over the high carrier density in the films, might allow the production of graphene with significantly larger domains and higher carrier mobility, even on wafers up to 12" in diameter.

This work was supported by the Defense Advanced Research Projects Agency (DARPA) under Contract No. FA8650-08-C-7838 [Carbon Electronics for RF Applications (CERA) program]. The authors are grateful to J. Ott, M. Freitag, D. Farmer, and S. J. Han for technical assistance and discussion, as well as to C. Y. Sung for the guidance and administration of the CERA project at IBM.  The views, opinions, and/or findings contained in this article are those of the authors and should not be interpreted as representing the official views or policies, either expressed or implied, of DARPA or the Department of Defense.

**Captions**

Figure 1: X-ray diffraction and TEM data from a 2" SiC on sapphire sample: (a) XRD (b) electron diffraction (c) cross-sectional TEM.

Figure 2: (a) Raman spectrum of MLG grown on a polycrystalline SiC on sapphire film showing a large defect band at 1358 cm$^{-1}$. (b) Cross-sectional TEM of the same film.

Figure 3 (color online): As grown SiC and MLG on sapphire (a) 10 μm x 10 μm topographic AFM before graphenization, (b) AFM and (c) Raman after graphenization to MLG. Planarized SiC and MLG on sapphire (d) AFM before graphenization, (e) AFM and (f) Raman after the graphenization. (g) and (h) cross sectional TEM of graphene on planarized SiC on sapphire. Inset: LEED of a similar MLG sample showing diffuse first order graphene spots.

Figure 1:

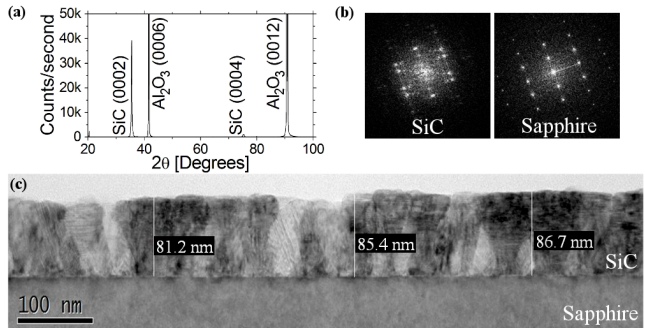

Figure 2:

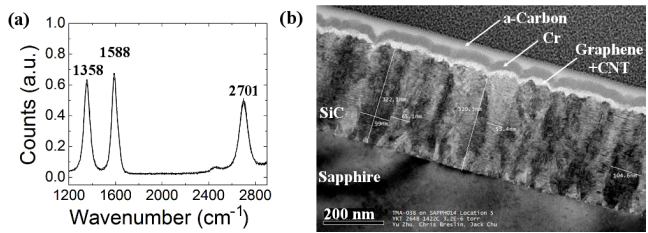

Figure 3:

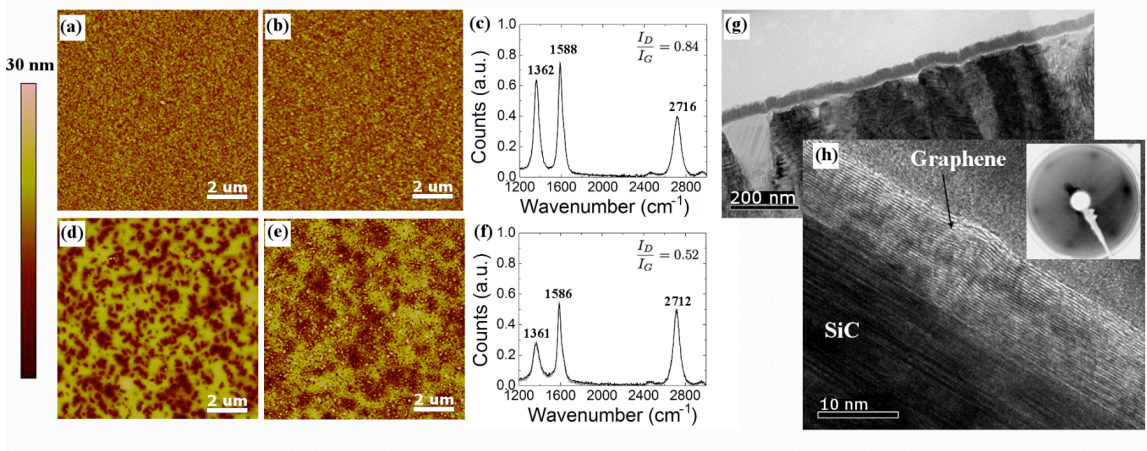